\documentclass[a4paper,usenatbib,onecolumn]{mn2e}
\usepackage{newtxtext,newtxmath}
\usepackage{lipsum}
\usepackage[T1]{fontenc}
\usepackage{ae,aecompl}
\usepackage{hyperref}
\usepackage{graphicx}	% Including figure files
\usepackage{amsmath}	% Advanced maths commands
\usepackage{amssymb}	% Extra maths symbols
\usepackage{bm}	% Extra maths symbols
\usepackage{caption}
\usepackage{multirow}
\usepackage{multicol}
\usepackage{cuted}
\usepackage{mathtools}
\usepackage{longtable}
\usepackage{rotating}

\title[ Was Proxima captured by alpha Centauri A and B]{Was Proxima captured by alpha Centauri A and B?}
\author[F. Feng et al.]
{F. Feng$^{1}$\thanks{E-mail: f.feng@herts.ac.uk or fengfabo@gmail.com}, H. R. A.  Jones$^{1}$\\
$^{1}$Centre for Astrophysics Research, University of Hertfordshire, College 
Lane, AL10 9AB, Hatfield, UK
}
\date{\today}

\begin{document}
\maketitle

\begin{abstract}
  The nearest stellar system consists of the stars Proxima, alpha Centauri A and B and at least one planet Proxima b. The habitability of Proxima b and any other planets are likely to be significantly influenced by the orbital evolution of the system. To study the dynamical evolution of the system, we simulate the motions of Proxima and alpha Centauri A and B due to the perturbations from the Galactic tide and stellar encounters in a Monte Carlo fashion. From 100 clones, we find that 74\% orbits of Proxima Centauri are bound to alpha Centauri A and B while 17\% and 9\% orbits become unbound in the simulations over the past and future 5\,Gyr. If the system migrated outward in the Milky Way to its current location, more than 50\% of clones could become unstable in backward simulations. The ratio of unstable clones increases with the simulation time scale and encounter rate. This provides some evidence for a capture scenario for the formation of the current triple system. Despite large uncertainties, the metallicity difference between Proxima and alpha Centauri A and B is also suggestive of their different origin. Nonetheless, further improvements in the available data and models will be necessary for a reliable assessment of the history of the Proxima-alpha Centauri system and its impact on the habitability of Proxima b.
\end{abstract}
%%%%%%%%%%%%%%%%%%%%%%%%%%%%%%%%%%%%%%%%%%%%%%%%%%%%%%%%%%%%%%%%%%%%%%%%%%%%
\section{Introduction}     \label{sec:introduction}
Since the discovery of Proxima Centauri b, a planet with a mass of 1.3\,$M_\oplus$ orbiting Proxima \citep{innes15} with a period of 11.2\,day \citep{anglada16}, intensive studies have been performed to investigate its habitability. Various scenarios for the formation of Proxima b have been proposed \citep{alibert16,coleman17} and its habitability has been comprehensively studied \citep{barnes16,ribas16}. The association between Proxima and alpha Centauri A and B have been frequently discussed in previous studies \citep{voute17,wertheimer06,matvienko14,pourbaix16}. Proxima, a late type M dwarf with a mass of 0.1221$\pm$0.022 $M_\odot$ \citep{mann15}, has been confirmed to be bound to alpha Centauri A (1.1055$\pm$0.0039\,$M_\odot$, \citealt{kervella16b}) and B (0.9373$\pm$0.0033\,$M_\odot$, \citealt{kervella16b}) by the recent use of the HARPS radial velocity data to constrain its orbit \citep{kervella17}. This triple system is composed of a close binary and a wide companion, which is a typical configuration for triples \citep{tokovinin06}. Proxima has a semi-major axis of 8.7$^{+0.7}_{-0.4}$\,kau, eccentricity of $0.50^{+0.08}_{-0.09}$, orbital period of 547$^{+66}_{-40}$\,kyr, and inclination of $107.6^{+1.8}_{-2.0}$\,deg with respect to alpha Centauri A and B \citep{kervella17}. 

According to \cite{barnes16}, the tidal force from alpha Centauri A and B is important for our understanding of the evolution and habitability of Proxima b. However, they did not account for the anisotropy in the velocity space of stellar encounters with respect to the primary star \citep{feng14}. They probably over-estimate the radial migration of the Sun and thus assume the formation of alpha Centauri A and B at a Galactocentric distance from 1.5\,kpc to 4.5\,kpc, which is inconsistent with recent studies of the Solar motions in the Galaxy accounting for the uncertainties of the Sun's motion (e.g. \citealt{feng13} and \citealt{martinez14}). In addition, they did not use the new data from \cite{kervella17} and thus adopt arbitrary initial conditions for the Proxima-alpha Centauri system. 

Although \cite{kervella17} find a $>10$\,Gyr stability of the Proxima and alpha Centauri system, the tidal radius \citep{jiang10} they have adopted is not reliable due to an underestimation of stellar density and unrealistic assumption of isotropic encounter velocities \citep{feng14}. Considering these limitations, we assess the stability of the Proxima-alpha Centauri system by simulating the motions of this system under the perturbations from the Galactic tide and stellar encounters. We will derive the initial conditions of alpha Centauri A and B and Proxima from the data in \cite{kervella17} and adopt a comprehensive encounter model according to \cite{feng14}. 

The paper is structured as follows. In section \ref{sec:method}, we describe the numerical method and the models for encounters and the Galactic tide. We report the results in section \ref{sec:results} and conclude and discuss in section \ref{sec:conclusion}. 

\section{Method}\label{sec:method}
We derive the initial conditions of Proxima and alpha Centauri in the Galactocentric reference frame from the data in \cite{kervella17}. The initial conditions of the Sun and the model of the Galactic potential are the same as in \cite{feng14}. We simulate the motions of clones under perturbations from the Galactic tide and stellar encounters using the Bulirsch-Stoer method \citep{bulirsch64} implemented in the {\small pracma} package of {\small R}. According to our tests, this integrator conserves the orbital energy and angular momentum with a relative numerical error down to $10^{-8}$ over 1\,Gyr with a time step of 0.01\,Myr. We also define a clone of the Proxima-alpha Centauri system as unstable if the eccentricity $e$ of Proxima with respect to alpha Centauri A and B is larger than one. Since some clones which become eccentric may be perturbed back to a bound state, a clone is considered as unstable only if it never comes back onto a stable orbit (i.e. $e<1$) within the simulation time span. 

Since the orbital period of Proxima around alpha Centauri is around 0.55\,Myr which is much longer than the interaction time scale of about 0.02\,Myr between encounters and the binary, assuming an encounter velocity of 50\,km/s according to \cite{feng14} and periapsis of 1\,pc. Thus we can apply the so-called ``impulse approximation'' \citep{rickman76} to calculate the perturbation of velocity from an encounter, which is
\begin{equation}
  \Delta{\vec{v}_*}=\frac{2GM_{\rm enc}}{v_{\rm enc}d_{\rm enc}}\vec{e}_{\rm enc}, 
  \end{equation}
  where $\vec{v}_*$ is the impulse gained by the target star, $G$ is the gravitational constant, $M_{\rm enc}$ is the encounter mass, $d_{\rm enc}$ is the impact parameter or periapsis, and $\vec{e}_{\rm enc}$ is the direction of periapsis. For every 1\,Myr, we randomly generate stellar encounters according to the model in \cite{feng14} and calculate the velocity kick for alpha Centauri A and B and Proxima. 

The main uncertain parameters in the above encounter model is the encounter rate. In \cite{feng17c}, we find the encounter rate for the Solar System is larger than 15 per Myr for encounters with perihelia less than 1\,pc, which is consistent with the value of 20 per Myr given by Bailer-Jones (2017) based on studies of TGAS encounter candidates. However, this value may be underestimated because GL 710 and the Scholtz's star are found to pass the Solar system at 0.06\,pc and 0.25\,pc about 1.35\,Myr from the present and 0.07\,Myr ago, respectively \citep{berski16,mamajek15}. This is roughly equivalent to one encounter passing the Sun within 0.1\,pc per Myr, although the sample of encounters are still incomplete \citep{feng15thesis,feng17c}. In other words, there are probably around 100 encounters passing the Sun within 1\,pc every Myr. 

We confirm this estimation using the formula of encounter rate which is
\begin{equation}
F=n\sigma\bar{v}_{\mathrm{enc}},
\label{eqn:flux}
\end{equation}
where $\sigma$, $\bar{v}_{\mathrm{enc}}$, $n$ are the cross section, velocity of the encounter in the helio-static frame and local stellar number density. According to \cite{binney08_book}, the cross section $\sigma$ is
\begin{equation}
  \sigma = \pi D^2\left(1+\frac{2G(M_{\mathrm{enc}}+M_\odot)}{Dv_{\mathrm{enc}}^2}\right), 
\end{equation}
where $D$ is the maximum perihelion or impact parameter of encounters, 
$M_\odot$ and $M_{\mathrm{enc}}$ are the masses of the Sun and the
encounters respectively, and $G$ is the gravitational constant. If $D$
is set to 1\,pc, the second term in the bracket is far less than
one. Thus the above cross section is approximately $\pi D^2$. 

The local number density is 0.098\,pc$^{-3}$ based on the sample of 211 stars within 8\,pc (excluding brown dwarfs) from the Sun collected by \cite{kirkpatrick12}. However, this number density is a lower limit of the real stellar number density due to the incompleteness of dwarf stars and other faint stars whose astrometry is unknown or poorly known. This is evident from the new discoveries of M, L, T and Y dwarfs (e.g., \citealt{finch14, finch16, kirkpatrick16}). The stellar density of local mid-plane is about 0.1$M_\odot$\,pc$^{-3}$ according to \cite{pham97,holmberg00,korchagin03,soubiran03}. The mean stellar mass is about 0.4\,$M_\odot$ based on the mass of 129 stars within 12 light years (or 6.44\,pc) from the Sun\footnote{\url{http://www.johnstonsarchive.net/astro/nearstar.html}}. Thus the stellar number density local to the Sun would be 0.25\,pc$^{-3}$. Assuming $\bar{v}_{\rm enc}=75$\,km/s according to \cite{feng17c}, the flux of encounters with $d_{\rm enc}<1$\,pc is about 60 per Myr.

  If we include brown dwarfs and other low-mass objects into our encounter model, the encounter rate could be higher. Compared with \cite{kirkpatrick12}'s sample, there are around 6 new brown dwarfs discovered within a heliocentric distance of 6.5\,pc \citep{bihain16}. In addition, the non-uniform distribution of nearby brown dwarfs indicates a bias in the observations of brown dwarfs \citep{bihain16}. This anisotropy may disappear due to new discoveries which would increase the brown-dwarf-to-star ratio and not change the local stellar number density much if the ratio is around 1 \citep{chabrier03}. Moreover, given ongoing new discoveries the current list of the nearest M dwarfs are probably not complete (e.g., \citealt{scholz13}). Considering all these factors, we set a default encounter rate of 80 per Myr and use the encounter model in \cite{rickman08} to generate encounters. We will also investigate the importance of the encounter rate for a sensitivity test in the next section.

Therefore we will simulate 320 encounters of alpha Centauri A and B with impact parameter $d_{\rm enc}$ less than 2\,pc per Myr, assuming that the local number density of alpha Centauri A and B is the same as the Solar System. The number of encounters per Myr is proportional to the local stellar density and thus is modulated during the migration of alpha Centauri A and B in the Galaxy. As the encounter velocity is calculated with respect to alpha Centauri and thus would not be isotropic, the mean directions of encounters would also be modulated due to a change of the peculiar motion of alpha Centauri A and B. 

We also follow \cite{feng15b} to predict the strength of an encounter using 
\begin{equation}
  g=\frac{M_{\rm enc}}{d^2_{\rm enc}v_{\rm enc}}~,
  \label{eqn:g}
\end{equation}
where $M_{\rm enc}$ is the mass of encounter. In this work, the impact parameter $d_{\rm enc}$ is calculated in the frame centered at the barycentre of Proxima and alpha Centauri.

\section{Results}\label{sec:results}
\begin{figure}
  \centering
  \includegraphics[scale=0.8]{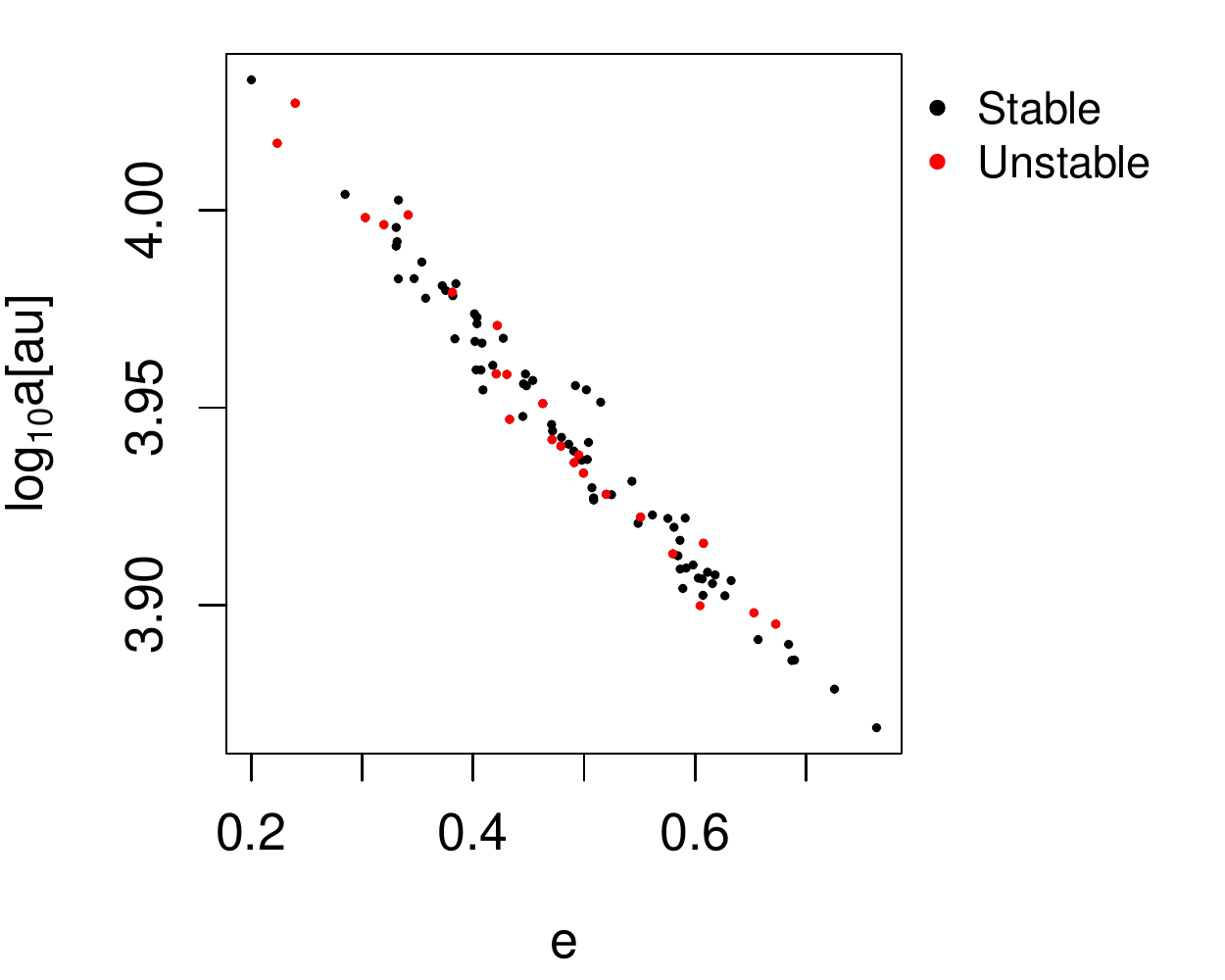}
  \caption{Distribution of initial eccentricity and semi-major axis of 100 clones of Proxima. The colours encode the final stage of their evolution in the simulations from -5\,Gyr to +5\,Gyr. The current encounter rate is set to be 80 encounters with $d_{\rm enc}<1$\,pc every Myr. }
  \label{fig:ic}
\end{figure}

We generate 100 clones of Proxima and alpha Centauri according to the data uncertainty given by \cite{kervella17}. We simulate the orbits of these clones under the perturbations from the Galactic tide and stellar encounters from -5\,Gyr to +5\,Gyr appropriate to the age for alpha Centauri A and B (e.g., 4.7-5.2\,Gyr, \citealt{bazot16}). The same sample of encounters are applied to different clones. We then show the distribution of initial eccentricity $e$ and semi-major axis $a$ of these clones with respect to the barycentre of alpha Centauri A and B in Fig. \ref{fig:ic}. We find 17\% and 9\% clones of Proxima were/will be ejected from the system, respectively, while the other clones experience strong orbital variation. Since the age of Proxima is probably older than 5\,Gyr \citep{eggenberger04}, we simulate another 100 clones of Proxima from -7\,Gyr to 7\,Gyr and find 23\% and 15\% clones of Proxima were/will be ejected from the system. Since the encounter sample is the same for all clones and some clones are ejected while others are not, the final status of clones are sensitive to their initial conditions. In other words, some clones have trajectories which are close to strong encounters while other clones are not. We also vary the encounter rate $F$ (see Eqn. \ref{eqn:flux}) and find an ejection ratio of 11\%, 31\%, and 54\% for simulations with the rate of encounters with $d_{\rm enc}<1$\,pc in units of Myr$^{-1}$, $F = $40, 60 and 100 from -7\,Gyr to 7\,Gyr, respectively. Based on a linear fit of the ejection ratio for these simulations, we find that the percentage of unstable clones $r$ is approximately $r\sim0.5F$. Thus the stability of the Centauri system is sensitive to the encounter rate. 

In Figure \ref{fig:ic}, we see that the clones which are ejected seem to be randomly scattered in the $e-a$ distribution. This distribution indicates a high correlation between $a$ and $e$, which approximately follows $a\propto10^{-e}$.  This is probably due to the fact that eccentricity or angular momentum is less sensitive to astrometric uncertainties than semi-major axis or orbital energy. The ejected clones follows a similar distribution to the stable ones, indicating that the ejection ratio is more sensitive to eccentricity than to semi-major axis. Thus a more precise measurements of orbital eccentricity is essential to improve the reconstruction of Proxima's dynamical history.

\begin{figure}
  \centering
  \includegraphics[scale=0.6]{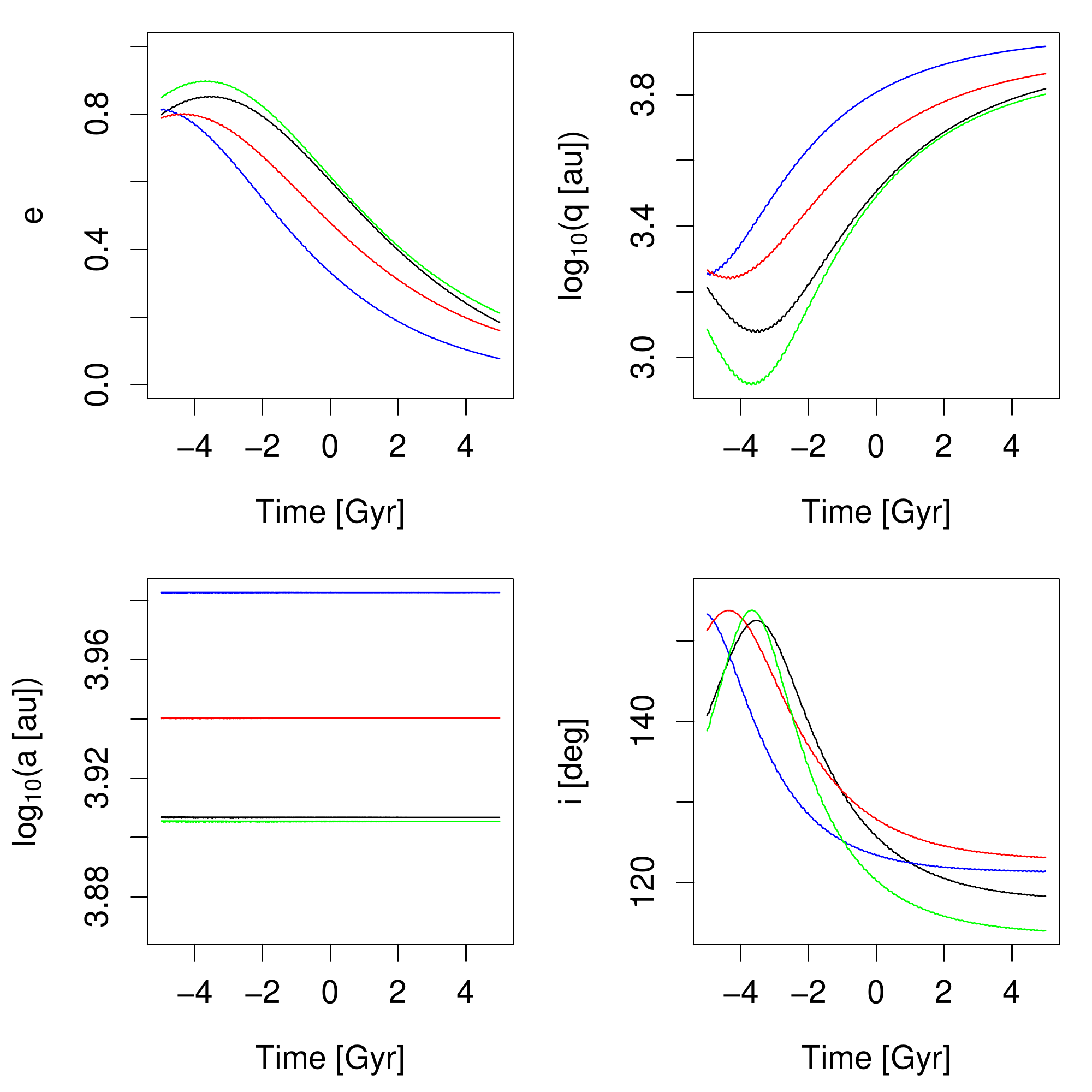}
  \caption{Variation of eccentricity $e$, periastron $q$, semi-major axis $a$, and inclination $i$, of Proxima under the perturbations from the Galactic tide based on simulations from -5\,Gyr to +5\,Gyr. The black and blue lines denote the orbital elements of clones which were/will be ejected through combined perturbations of stellar encounters and the Galactic tide. The red and green lines denote the orbits of stable clones. }
  \label{fig:tide}
\end{figure}

\begin{figure}
  \centering
  \includegraphics[scale=0.6]{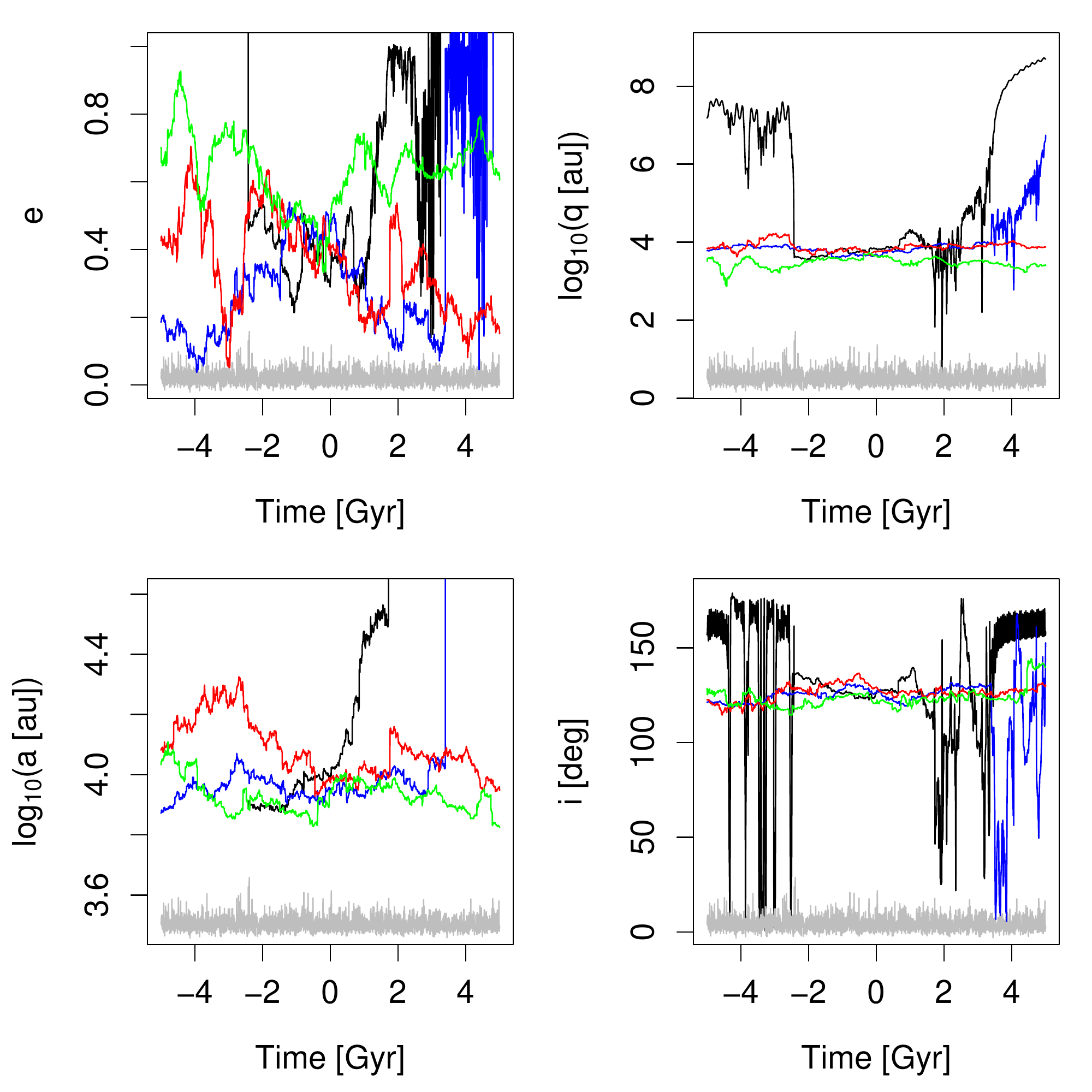}
  \caption{Similar to Fig. \ref{fig:tide}, but for orbital elements of Proxima under the combined perturbations from the Galactic tide and stellar encounters. The grey lines at the bottom shows the strength of encounters characterized by $g$ (see Eqn. \ref{eqn:g}), which is smoothed with a 1\,Myr time bin. }
  \label{fig:te}
\end{figure}

To see the effect of encounters, we select two stable and two unstable clones and show the evolution of their orbital elements under the perturbations of the Galactic tide alone and of both the Galactic tide and encounters in Fig. \ref{fig:tide} and in Fig. \ref{fig:te}, respectively. Without stellar encounters, the semi-major axes or orbital energy of clones are almost constant over time. The eccentricity, periastron and inclination vary gradually and continuously in the whole simulations, and the evolution of orbital elements is sensitive to the argument of periastron \citep{veras12}. However, if encounters are included into the simulations, all orbital elements vary significantly and stochastically due to the random impulses from encounters. Apart from the orbital elements shown in Fig. \ref{fig:te}, we also find significant variation in the argument of periastron $\omega$ and relatively weaker variation in the longitude of ascending node $\Omega$, indicating a more sensitive response of $\omega$ to encounter perturbations than $\Omega$.

In Fig. \ref{fig:te}, we also see a strong connection between the jump of the orbital elements and the peaks in the bottom grey lines. That means the strength of the perturbation imposed by an encounter is well characterized by the proxy of $g$ defined in Equation \ref{eqn:g}. This is consistent with the conclusion based on the study of the Oort cloud by \cite{feng15b} and the analytical expression of encounter-induced eccentricity change derived by \cite{heggi96}. 

The orbits of clones are especially sensitive to strong encounters, which are slow, massive and extremely close to the clones. This is evident from the abrupt change of semi-major axis caused by strong encounters. However, strong encounters only play a role to trigger the escape of clones from the system while weak encounters cumulatively randomize the clones' orbits and thus pave the way for escape. In other words, the orbital variation is more like a stochastic process than a deterministic process. Hence an analytical escape radius/zone based on the occurrence rate of strong encounters \citep{veras13,zwart15} is probably not reliable for stability analysis for multiples, especially for wide multiples like the Proxima-alpha Centauri system. But a statistical analog to the escape radius can be derived based on simulations of a large sample of wide multiples covering larger parameter space, which is beyond the scope of this work.

\section{Discussion and conclusion}\label{sec:conclusion}
While Proxima Centauri has been of particular recent interest due to the discovery of Proxima b, the alpha Centauri system has long been studied for many reasons by a variety of different techniques. Of particular interest to us is the extent to which these studies support a common origin for all three stars. Given the benchmark status of the system there are a number of diagnostics available. 

In this work, we investigate the capture scenario by simulating the Proxima-alpha Centauri system over the past and future 5 and 7\,Gyr and find a considerable percentage of Proxima's clones become unstable in backward simulations. Although most N-body simulations are not strictly time-reversible due to cumulative numerical errors \citep{rein17}, the simulations in this work are time reversible in a statistical sense. The fact that some clones of Proxima become unstable by looking backward is equivalent to them becoming stabilized by looking forward. Thus our simulations not only provide evidence against the {\it in situ} scenario but also for the capture scenario. In addition, many disc stars are likely to have migrated from near the Galactic centre to their current locations \citep{sellwood02}. If the Proxima-alpha Centauri system formed near the Galactic centre, it would experience more intensive perturbations from the Galactic tide and encounters and more Proxima clones would become unbound in backward simulations, and thus the capture scenario would be more favored. For example, according to \cite{barnes16}, the Proxima-alpha Centauri system has migrated outward at least 3.5\,kpc, corresponding to a stellar density five times higher at the formation region than at the Sun's current location. Thus the encounter rate would also be at least five times higher during the early evolution of the system, leading to more than 50\% ejection percentage assuming $r\sim0.5F$ and a linear radial migration.

Moreover, the capture of a field star is not rare if the rate of encounters with $d_{\rm enc}<1$\,pc is 80 every Myr, equivalent to about one encounter with $d_{\rm enc}<0.1$\,pc per Myr. There are more than 0.1\% encounters moving extremely slowly with respect to the reference star, with a relative velocity less than 1\,km/s \citep{feng17c}. Hence a Sun-like system would encounter a few extremely close and slow encounters after the dissolution of its birth cluster. These encounters could be captured by the system and become stabilized through the perturbations from the Galactic tide and stellar encounters, as seen in the simulations of Proxima clones. Moreover, the eccentricity evolution in tide-only simulations shown in Fig. \ref{fig:tide} indicates that Proxima was on a highly eccentric orbit about 5\,Gyr ago, probably corresponding to a unstable stage after capture. This capture scenario provides another channel for the formation of multiples, in addition to the dynamical unfolding scenario \citep{reipurth12}. If the capture scenario is plausible for Proxima, it would be more plausible for the formation of wider multiples. A comparative study of the capture and {\it in situ} scenarios (e.g. \citealt{kouwenhoven10, moeckel11}) would help to infer the formation channels from the orbital characteristics of the current sample of wide binaries (e.g. \citealt{caballero07,price-whelan17}).

The dynamical evolution of the Proxima-alpha Centauri triple is sensitive to its encounter history. However, the current sample of encounters of alpha Centauri A and B is too small for a realistic reconstruction of the encounter history even within 10\,Myr \citep{feng17c}. With the upcoming Gaia data releases, we expect to identify more encounters within the past 1\,Gyr \citep{feng15thesis}. This would provide a realistic sample for the reconstruction of the dynamical history of Proxima and thus for the assessment of the dynamical habitability of Proxima b and other potential planets in the Proxima-alpha Centauri system.

Considering that many clones are ejected in the 5\,Gyr simulations, the dissolution time of $>10$\,Gyr introduced by \cite{jiang10} is probably not a reliable metric to estimate the long-term stability of wide binaries. This is probably due to their underestimation of the encounter rate based on a stellar density of 0.05\,pc$^{-3}$. In addition, they assume isotropic encounter velocities or equivalently that the Sun is static with respect to the local standard of rest, which is unrealistic \citep{feng14}. A revised version of tidal radius based on realistic encounter model is needed to estimate the stability of wide binaries. 

The tidal force caused by the quadratic potential of alpha Centauri A and B is not accounted for in our simulations. This force becomes significant when the periastron of Proxima is less than 2000\,au \citep{barnes16}. In our simulations, few stable orbits of clones have periastrons below this limit while the unstable orbits tend to have lower periastron before being ejected. Despite this most unstable clones would become unbound at their apsides due to stronger and longer perturbations from encounters and the Galactic tide. In other words, most unstable clones will be kicked out of the system before arriving at their periapses. Thus the quadratic potential of A and B is not important for the estimation of the ejection ratio. But it is probably essential for a comprehensive study of the dynamical habitability of Proxima b and other potential planets in the alpha Centauri A and B since the members in a hierarchical multiple system might influence the secular evolution of each other according to studies of hierarchical multiples \citep{hamers15}.

In addition to dynamical studies, coevality of the components in the Proxima-alpha Centauri system can be judged from the consistency of their evolutionary history as determined by different isochrone fits to their luminosity and mass and the similarity of their metallicity as judged by spectral synthesis. For alpha Cen A and B, there is a long history of studies which converge to find them as having a significantly metal-rich composition and solar-type age. For example, \cite{Jofre15} use eight different methods to determine abundances to find [Fe/H] = 0.26$\pm$0.08 and 0.22$\pm$0.10 respectively and do not find any particularly peculiarities relative to a differential analysis with respect to the Sun other than for Mn which is expected due to its sensitivity to micro-turbulence. In terms of age, \cite{bazot16} consider the range of available astroseismic data for alpha Cen A with applicable models in the range of 4.7-5.2\,Gyr.

Derivations of age and metallicity for Proxima are substantially more uncertain since it is a late type M dwarf which are relatively less well-studied. The question of the evolutionary history of the alpha Centauri system and in particularly the metallicity for Proxima has been recently addressed by \cite{beech17}. They find a best match for Proxima to have a metallicity of [Fe/H]$=-0.5$ or by constraining a simultaneous fit with the values of mass and radius used by \cite{kervella17} a metallicity of [Fe/H]$=-0.3$. However, there are a number of uncertainties involved and a metallicity as high as solar is allowed within their one-sigma error bars when considering a mass-radius relationship based on absolute K magnitude. There are a number of recent derivations of spectroscopic metallicities for Proxima showing a spread around a solar-like value, e.g., -0.07$\pm$0.14 \citep{passegger16}, 0.05$\pm$0.20 \citep{kervella16}. For hotter stars ages maybe reliably determined by astroseismic and activity measurements. However, late type M dwarfs like Proxima lie in a rather poorly calibrated regime. Proxima presents both a high flare rate (e.g., \citealt{davenport16}) and slow period (e.g., \citealt{collins16}). The data from Sloan Digital Sky Survey indicate that late type M5 and M6 dwarfs have activity lifetimes of 7$\pm$0.5\,Gyr \citep{west08}.  So while the metallicity of Proxima (and of all late type M dwarfs) is relatively poorly constrained recent literature results are suggestive of a metallicity difference between Proxima and alpha Centauri A and B which would be consistent with different formation histories. 

As concluded in \cite{barnes16}, alpha Centauri A and B could be close enough to destabilize the orbit of Proxima b. However, if Proxima was captured by alpha Centauri, Proxima b may only be influenced by alpha Centauri A and B for a relatively short period of time, and thus was probably orbiting in the current habitable zone for a long time. In summary, a comprehensive study of the dynamical history and composition of Proxima and alpha Centauri A and B is crucial for a complete understanding of their formation and evolution and the habitability of their planets. 

\section*{Acknowledgement}
This work is supported by the Science and Technology Facilities Council (ST/M001008/1). We used the ESO Science Archive Facility to collect radial velocity data. We thank Coryn Bailer-Jones, Eric Mamajek, Martin Beech, and Jose Caballero for helpful discussions, which significantly improve the manuscript. We also thank the anonymous referee for valuable comments. 
\bibliographystyle{aasjournal}
\bibliography{nm}
% \end{CJK*}
\end{document}